\documentclass[twocolumn]{article}
\usepackage[utf8x]{inputenc}
\usepackage[english]{babel}
\usepackage{authblk}
\usepackage{geometry}
\usepackage{amsmath}
\usepackage{amssymb}
\usepackage{natbib} 
\usepackage{graphicx}
\usepackage{xcolor}
\usepackage{abstract}
\bibpunct{(}{)}{,}{a}{}{,}
\usepackage[colorlinks=true,linkcolor=blue,citecolor=blue]{hyperref}

\advance\topmargin 1cm

\title{Cosmological observational tests in the JWST Era.~I: angular size --- redshift}

\author[1,2]{A.~A.~Raikov}
\author[3,4]{V.~V.~Tsymbal}
\author[5,6]{N.~Yu.~Lovyagin}

\affil[1]{Saint Petersburg Branch of the Special Astrophysical Observatory of RAS, 196140, 65 Pulkovskoe Shosse, Saint Petersburg, Russia}
\affil[2]{Main Astronomical Observatory of RAS, 196140, 65 Pulkovskoe Shosse, Saint Petersburg, Russia}
\affil[3]{Institute of Astronomy of RAS, 119017, 48 Pyatnitskaya St., Moscow, Russia}
\affil[4]{Special Astrophysical Observatory of RAS, 369167, Nizhnij Arkhyz, Russia}
\affil[5]{Saint Petersburg State University, 199034, Universitetskaya Emb., 7/9, Saint Petersburg, Russia, e-mail: n.lovyagin@spbu.ru}
\affil[6]{Saint Petersburg State Marine Technical University, 190008, Lotsmanskaya St. 3, Saint Petersburg, Russia}

\date{}

\begin{document}

\twocolumn[
\maketitle 
\begin{onecolabstract}
		This study is devoted to the cosmological ``angular size --- redshift'' test. An analysis is performed of the angular and linear sizes of galaxies from the new ASTRODEEP-JWST catalogue, which contains over 500\,000 objects at high redshifts (up to $\sim 20$ photometrically determined and up to $\sim 14$ spectroscopically determined). For the calculations, 6\,860 galaxies with reliably determined spectroscopic redshifts and 319\,771 galaxies with known photometric redshifts were used. The linear sizes of galaxies were computed within the framework of two cosmological models --- the standard ($\Lambda$CDM) model and one of the static models (the so-called ``tired light'' model). We have shown that within the framework of the $\Lambda$CDM model, a significant evolution of the linear sizes of galaxies is observed, with the rate of the evolution closely matching the rate of the cosmic expansion. In contrast, in the static model, the characteristic linear sizes of galaxies exhibit almost no evolution with increasing $z$.
		
\vspace*{0.7em}
		
\vspace*{0.7em}

\vspace*{0.5em}
\noindent {\slshape This is a preprint of a manuscript published in the \textit{Astrophysical Bulletin, 2025, Vol. 80, No. 3, pp. 337--347}.

}
\vspace*{1cm}
	\end{onecolabstract}
]

\section{Introduction}

Observational tests are a key tool for determining the parameters and evaluating the viability of cosmological models. Their importance is emphasized, for example, in the work of \citet{geller1972test}. With the establishment of the $\Lambda$CDM model as the standard cosmological model and the advent of ``precision cosmology'' based on observations of the cosmic microwave background by the WMAP \citep{spergel2003first} and Planck \citep{aghanim2020planck} space observatories, observational tests---especially those based on data from the local Universe---are often given less attention and considered of secondary importance in the field.

However, new observational data have revealed a wide range of issues within the standard cosmological model. In particular, the value of the Hubble constant \citep{di2021realm,kamionkowski2023hubble}, as determined from observations of the local Universe, and the value of the optical depth of reionization (i.e., the number of ionizing photons in the early stages of cosmic evolution), as inferred from JWST observations \citep{melia2024cosmic,munoz2024reionization}, show significant tension with the results of Planck ``precision cosmology''. A number of problems remain unresolved, such as the lack of a model capable of explaining the formation of the galaxies observed at high $z$ within the time corresponding to their ages as inferred under the $\Lambda$CDM model (as discussed, in particular, in \cite{dolgov2018mass}). These issues are further highlighted by the JWST observations. In this context, the need to perform cosmological tests becomes relevant again, and the reconsideration of cosmological theories alternative to the standard model may once again prove inevitable, as was previously emphasized in \cite{orlov2016cosmological}.

In this work, we revisit the ``angular size---redshift'' test. This test is based on the fact that the functional dependence of the angular size of a standard ruler on redshift differs significantly between the standard cosmological model (an expanding Universe) and the static cosmological model---by a factor of the order of $(1+z)^1$.

The test was originally proposed by \cite{hoyle1959relation}. A comparison of the test results for static models and models with expanding space was presented by \cite{geller1972test}. Studies of the sizes of galaxies within the standard $\Lambda$CDM cosmological model were presented by \cite{allen2017size,holwerda2020sizes}, who drawn a conclusion was drawn regarding the significant evolution of the linear sizes of galaxies. A comparative analysis of the theoretical predictions for the angular size---redshift relation in various cosmological models, including both static models and models with space expansion, was given by \cite{lopez2010angular}; it was shown that, as of the observational data available at the time (2010), the model with a linear Hubble law provided the best agreement.

In the work by \cite{lovyagin2022cosmological}, the ``angular size---redshift'' test incorporated the first observations from the JWST telescope. The study was conducted using a relatively small set of observational data and relied on the photometric redshifts. It was shown that the initial JWST observations are in better agreement with the static cosmological model than with the expanding one, provided that the evolution of the linear sizes of galaxies is neglected. Authors used the ``tired light'' static model as a comparison to the standard $\Lambda$CDM model. In \cite{laviolette2021expanding}, it was also demonstrated that the static ``tired light'' model provides a better fit to the observational data within the framework of this test.

Recently, the unique ASTRODEEP-JWST catalogue \citep{merlin2024astrodeep}, containing more than 500,000 galaxies at high redshifts (up to $\lesssim 20$), has become available, opening previously inaccessible opportunities for testing cosmological theories. In the present study, we analyse the ``angular size---redshift'' test of galaxies on the basis of data provided by this catalogue.

For the analysis, we selected 6,860 galaxies from the catalogue with reliably determined spectroscopic redshifts and 319,771 galaxies with known photometric redshifts. The angular and linear sizes of galaxies were analysed within the theoretical predictions of two cosmological models---the standard ($\Lambda$CDM) model and one of the static models.

As the static model, we chose the so-called ``tired light'' model due to its simplicity and wide recognition. We do not claim here that the ``tired light'' model is necessarily correct. Our goal is to demonstrate the difference between static Universe models and models in which cosmological redshift is a consequence of space expansion.

To date, a fairly broad discussion exists in the scientific literature regarding the problems and paradoxes of the $\Lambda$CDM model, in particular the difficulties in explaining the rate of galaxy formation, as well as arguments supporting the possibility of alternative explanations of observations within static and other cosmological models. Some examples are given in \cite{dolgov2018mass,lopez2022alternative,peebles2022anomalies,perivolaropoulos2022challenges}

A detailed review and analysis of the full range of modern alternative cosmological models is a separate task and is beyond the scope of this article.

\section{Angular size --- redshift test}

The formulae presented in this section are adopted from \cite{hogg1999distance, lopez2010angular, laviolette2021expanding}.

The dependence of the angular size $\theta$ of an object on its linear size $D$ and cosmological redshift $z$ is expressed via the angular diameter distance $D_A(z)$:
\begin{equation}
	D_A(z)= \frac{D}{\theta(z)}.
\end{equation} 

The behaviour of $D_A(z)$ differs between cosmological models. In the standard ($\Lambda$CDM) cosmological model, it is given by
\begin{multline}
	D_A^{\Lambda{\rm CDM}}(z)=\frac{c}{H_0} \frac{1}{1+z} \times \\ \times \int\limits_0^z \frac{\mathrm dz'}{\sqrt{			\Omega_M (1+z')^3 + \Omega_\Lambda}}, 
	\label{eq:angular_diam_integr}
\\
	\mbox{where }\Omega_M + \Omega_\Lambda = 1.
\end{multline}

Assuming a linear Hubble law (the ``na\"\i ve Euclidean model of the Universe''),
\begin{equation}
	D_A^{Linear}(z)=\frac{c}{H_0} z.
\end{equation}  
In the static ``tired light'' model,
\begin{equation}
	D_A^{\rm TL}(z) = \frac{c}{H_0} \ln(1+z)\,. 
	\label{eq:tl_and_diam_dist}
\end{equation} 

Fig.~\ref{fig:z-theta} shows the theoretical angular size–redshift relations, along with observational data, for a galaxy of size $D=10$~kpc.

Unless otherwise specified, all calculations are performed using the following values for the parameters of the standard cosmological model: $H_0 = 70$~km/s/Mpc, $\Omega_\Lambda = 0.7$, $\Omega_M = 0.3$, and $\Omega_K = 0$.

\section{Data used}

The ASTRODEEP-JWST catalogue \citep{merlin2024astrodeep} contains information for 531,173 galaxies, including equatorial coordinates, rectangular coordinates in a fixed mosaic system, photometric parameters, radii enclosing 50\% of the galaxy's light (also referred to as ``effective radii''), morphological characteristics of each object, fluxes in 16 photometric bands covering the range of approximately $\sim\nolinebreak 4300$–$45000$~\AA, and the associated measurement errors. In addition, photometric redshifts were estimated for each object using four different methods, and spectroscopic redshifts are provided for galaxies for which spectroscopic observations were available. Each object is also assigned a quality flag ($flags$) indicating the reliability of the measured observational properties. The total sky coverage of this survey is approximately $\sim 0.2$ square degrees.

The main advantage of the ASTRODEEP-JWST catalogue is the homogeneity of the data it contains. The authors of the database combined existing photometric datasets from HST and JWST observations, reduced the images into a fixed mosaic system, and performed a unified recalibration of the data. Only after this step were the measured quantities included in the database.

For the purposes of this test, we initially chose to use galaxies with reliably determined spectroscopic redshifts. Although this limits the available sample size and the maximum redshift reached, the resulting dataset is free from the known uncertainties associated with photometric redshifts. Therefore, the test results based on spectroscopic redshifts are the most robust.

We also included only galaxies observed with JWST. All objects with poor-quality measurements, blended sources, those near the image edges, point-like sources, and other problematic cases were excluded (as a selection criterion, we used $flags < 80$). As a result, we selected 6,860 galaxies with reliably determined spectroscopic redshifts and 319,771 galaxies with (methodologically) reliably determined photometric redshifts. For these, angular sizes (defined as twice the effective radius) were extracted, and linear sizes (corresponding effective diameters) were computed within the two cosmological models under consideration.

\section{Catalogue analysis results}

\subsection{Angular size --- redshift diagram}

\begin{figure}
	\includegraphics[width=\hsize]{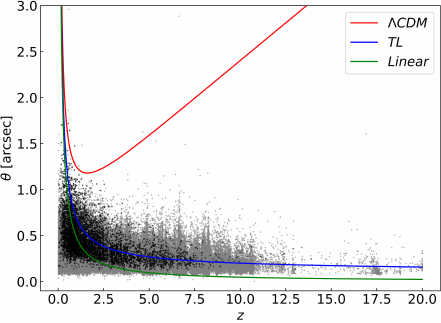}
    \caption{Angular size as a function of redshift. Observational data (galaxies) are shown as points: black for spectroscopic $z$, grey for photometric $z$. Theoretical curves for a linear size of $D = 10$~kpc are shown as solid lines for the $\Lambda$CDM model (red), the ``tired light'' model (blue), and the linear Hubble law (green). Angular sizes of galaxies were taken as twice the catalogue value of the effective radius.}
	\label{fig:z-theta}
\end{figure}

Fig.~\ref{fig:z-theta} shows the angular sizes of galaxies as a function of redshift, compared with theoretical angular size---redshift relations for a ``galaxy'' (standard ruler) with a linear size of $D = 10$~kpc, calculated for the $\Lambda$CDM cosmological model, the ``tired light'' model, and the linear Hubble law model.

Fig.~\ref{fig:z-theta} shows that, within the $\Lambda$CDM model, a significant rate of evolution in the linear sizes of galaxies over time is observed, whereas in the static ``tired light'' model, the dependence of the characteristic angular size of catalogue galaxies on redshift closely follows the theoretical angular size---redshift relation for a standard ruler.

\subsection{Linear size --- redshift diagram}

\begin{figure}
	\includegraphics[width=\hsize]{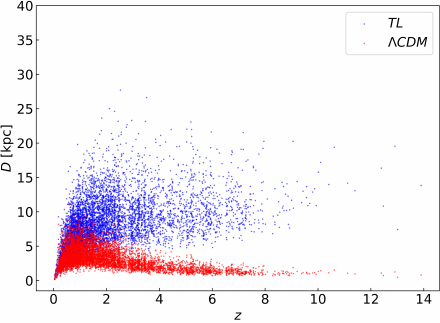}
	\includegraphics[width=\hsize]{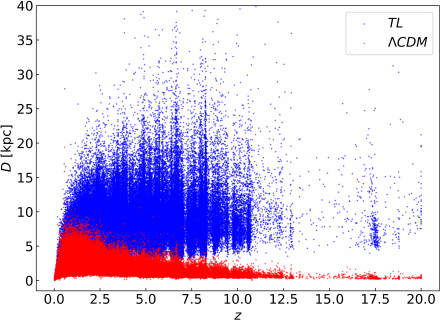}
	\caption{Linear sizes (effective diameters) of galaxies as a function of redshift. Top panel: only galaxies with the spectroscopic redshifts; bottom panel: galaxies with the photometric $z$. Red points correspond to calculations based on the $\Lambda$CDM model; blue points to those based on the ``tired light'' model.}
	\label{fig:z-size}
\end{figure}

Fig.~\ref{fig:z-size} presents the results of calculating the linear sizes of galaxies as a function of redshift, performed within the frameworks of the $\Lambda$CDM model and the ``tired light'' model.

It can be seen that, within the standard cosmological model, the most luminous galaxies at high redshift have typical sizes of $\sim 1$–2~kpc, whereas in the ``tired light'' model their sizes correspond to those of giant galaxies in the present epoch. The scatter of data points (black points in Fig.~\ref{fig:z-theta}) resembles the profile typically observed in wide-angle surveys of galaxies in the local Universe, caused by observational selection of bright objects (Malmquist bias). Thus, the linear sizes of galaxies calculated under the static ``tired light'' model show no evidence of evolution.

\section{Evolution of galaxy linear sizes}

\subsection{Analysis based on the scatter diagram}\label{rass}

Comparison of the angular size---redshift relation for galaxies with theoretical predictions (Fig.~\ref{fig:z-theta}), and the behaviour of linear galaxy sizes in the two models (Fig.~\ref{fig:z-size}), shows that in the absence of spatial expansion, no significant evolution in the linear sizes of galaxies with $z$ is observed. It can also be hypothesized that, within the framework of the standard $\Lambda$CDM cosmological model, the rate of evolution of galaxy sizes is of the same order as the expansion rate of the Universe---i.e., $\propto (1+z)^{-1}$.

To estimate the rate of evolution more accurately and to qualitatively assess the impact of observational selection on surface brightness and apparent magnitude, a series of galaxy subsamples was considered. Galaxies with redshifts lower than a given threshold $z_{min}$ were excluded (i.e., only those satisfying $z \geqslant z_{min}$ were included).

For each subsample, a linear regression was performed in logarithmic coordinates $\bigl(\lg(1+z), \lg D\bigr)$, using the least-squares method to determine the exponent $\alpha$ in the relation $D \propto (1+z)^{\alpha}$, where $D$ is the effective galaxy diameter. Since the catalogue does not provide errors for the effective radii, the standard (unweighted) matrix least-squares method was used. The formal uncertainty in $\alpha$ reflects the scatter in galaxy sizes in the diagram.

\begin{figure}
\includegraphics[width=\hsize]{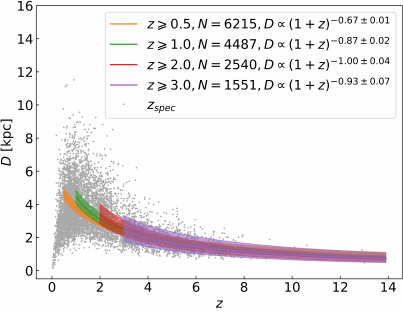}
\includegraphics[width=\hsize]{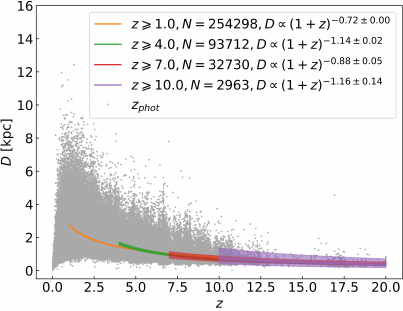}
\caption{Evolution of the linear sizes (effective diameters) of galaxies within the $\Lambda$CDM model. Data points represent individual galaxies; the curves show best-fit regression lines for subsamples with different minimum redshifts. Shaded regions indicate the $3\sigma$ uncertainty of the fit. Top panel: spectroscopic redshifts; bottom panel: photometric redshifts.}
\label{fig:z-Dasmpt}
	
\end{figure}

The results are presented in Table~\ref{tab:asmpt}, and the best-fit curves for selected subsamples are shown in Fig.~\ref{fig:z-Dasmpt}.

\begin{table}
\caption{Analysis of the galaxy size rate of evolution within the $\Lambda$CDM model. Galaxy subsamples were selected by applying a lower limit on redshift: $z \geqslant z_{min}$. The table shows the number of galaxies in each subsample ($N_{obj}$) and the best-fit power-law index $\alpha$ from the relation $D \propto (1+z)^{\alpha}$, along with its formal $3\sigma$ uncertainty. Subsamples based on photometric redshifts with low statistics and large uncertainties in $\alpha$ are shown in grey.}
\label{tab:asmpt}
\begin{tabular}{c||c|c||c|c}	
& \multicolumn{2}{c||}{Spectroscopic $z$} & \multicolumn{2}{c}{Photometric $z$} \\ \hline
$z_{min}$ & $n_{obj}$ & $\alpha \pm 3\sigma_\alpha$ & $n_{obj}$ & $\alpha \pm 3\sigma_\alpha$ \\ \hline
0.0 & 6860 & $-0.48 \pm 0.01$ & 319722 & $-0.13 \pm 0.00$ \\ 
0.5 & 6215 & $-0.67 \pm 0.01$ & 287303 & $-0.59 \pm 0.00$ \\ 
1.0 & 4487 & $-0.87 \pm 0.02$ & 254298 & $-0.72 \pm 0.00$ \\ 
1.5 & 3326 & $-0.97 \pm 0.03$ & 213249 & $-0.85 \pm 0.00$ \\ 
2.0 & 2540 & $-1.00 \pm 0.04$ & 176191 & $-0.97 \pm 0.01$ \\ 
2.5 & 1890 & $-0.95 \pm 0.06$ & 147935 & $-1.02 \pm 0.01$ \\ 
3.0 & 1551 & $-0.93 \pm 0.07$ & 130626 & $-1.09 \pm 0.01$ \\ 
3.5 & 1161 & $-0.88 \pm 0.10$ & 113110 & $-1.09 \pm 0.01$ \\ 
4.0 & 931 & $-0.86 \pm 0.13$ & 93712 & $-1.14 \pm 0.02$ \\ 
4.5 & 755 & $-0.86 \pm 0.17$ & 88065 & $-1.10 \pm 0.02$ \\ 
5.0 & 634 & $-0.82 \pm 0.20$ & 74201 & $-1.06 \pm 0.02$ \\ 
5.5 & 463 & $-0.76 \pm 0.24$ & 65118 & $-1.03 \pm 0.03$ \\ 
6.0 & \textcolor{gray}{301} & \textcolor{gray}{$-0.67 \pm 0.32$} & 53767 & $-0.83 \pm 0.03$ \\ 
6.5 & \textcolor{gray}{231} & \textcolor{gray}{$-0.74 \pm 0.38$} & 46063 & $-0.72 \pm 0.04$ \\ 
7.0 & \textcolor{gray}{142} & \textcolor{gray}{$-0.51 \pm 0.50$} & 32730 & $-0.88 \pm 0.05$ \\ 
7.5 & \textcolor{gray}{73} & \textcolor{gray}{$-0.61 \pm 0.72$} & 29535 & $-0.84 \pm 0.05$ \\ 
8.0 & \textcolor{gray}{47} & \textcolor{gray}{$-0.33 \pm 0.88$} & 24054 & $-0.73 \pm 0.06$ \\ 
8.5 & \textcolor{gray}{39} & \textcolor{gray}{$-0.60 \pm 0.97$} & 8597 & $-1.13 \pm 0.08$ \\ 
9.0 & \textcolor{gray}{24} & \textcolor{gray}{$-1.16 \pm 1.31$} & 5943 & $-1.18 \pm 0.10$ \\ 
9.5 & \textcolor{gray}{18} & \textcolor{gray}{$-1.17 \pm 1.65$} & 4300 & $-1.07 \pm 0.11$ \\ 
10.0 & \textcolor{gray}{11} & \textcolor{gray}{$-2.19 \pm 2.69$} & 2963 & $-1.16 \pm 0.14$ \\ 
10.5 & \textcolor{gray}{8} & \textcolor{gray}{$-2.46 \pm 4.93$} & 1971 & $-1.24 \pm 0.16$ \\ 
11.0 & \textcolor{gray}{7} & \textcolor{gray}{$-1.61 \pm 6.54$} & 1033 & $-1.56 \pm 0.25$ \\ 
\end{tabular}	
\end{table}

From the table and figure, it is evident that at low values of $z_{min}$, the estimated value of the exponent $\alpha$ is strongly affected by nearby small galaxies, which are underrepresented at higher redshifts due to observational selection effects. Starting from $z > 1$, the rate of evolution of galaxy sizes is approximately in the range $0.85$--$1.0$. The analysis based on the photometric redshifts yields similar results and extends the trend to higher redshifts.

\subsection{Analysis based on the mean galaxy size per bin}

To further investigate the rate of evolution, we performed bin-based statistics. The entire $z$ range was divided into 50 bins, where the mean linear diameters and their error were computed:
$$s = \sigma_D/\sqrt{n},\quad \mbox{where } \sigma^2_D = \frac1{n-1} \sum_{k=1}^{n} (\bar D - D_k)^2,$$
where $\bar D$ is the mean galaxy size in the bin, and $n$ is the number of galaxies in the bin. This formal error was used as the bin weight in the least squares method,
$W = 1/s^2$, which determines the formal uncertainty of the parameter $\alpha$ in the scaling $D \propto (1+z)^\alpha$.

\begin{figure}
	\includegraphics[width=\hsize]{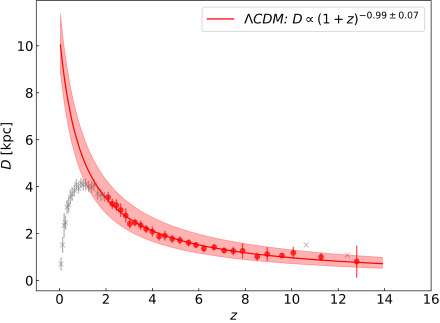}
	\includegraphics[width=\hsize]{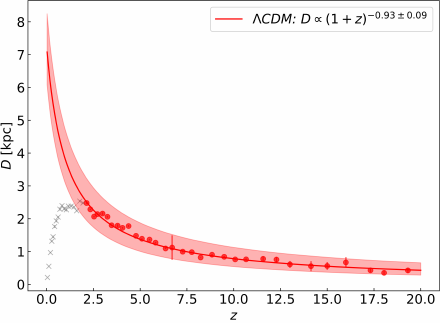}
	\caption{Evolution of galaxy linear sizes (effective diameters) in the $\Lambda$CDM model, statistics for 50 redshift bins. Points show the mean galaxy sizes in each bin with their standard errors of the mean (grey points denote bins excluded due to low redshift or insufficient number of objects). Lines indicate the best-fit curves; shaded areas show the $3\sigma$ formal fitting uncertainty. Top: spectroscopic redshifts, bottom: photometric redshifts.}	
	\label{fig:zbin}
\end{figure}

The bins were taken uniformly in linear coordinates, but the weighted least-squares fitting was applied to the values in the coordinates $\bigl(\lg(1+z), \lg D\bigr)$ with the corresponding transformation of the formal error $s_{\log} = 1 / \ln 10 \cdot s / \bar D$. Bins with low redshift ($z<2$) and bins containing too few galaxies ($n<2$) were excluded from the analysis. The results are shown in Fig.~\ref{fig:zbin}.

This analysis also shows a similar rate of galaxy evolution: $0.99 \pm 0.07$ (spectroscopic $z$) and $0.93 \pm 0.09$ (photometric $z$). The results for the spectroscopically and photometrically determined redshifts are indistinguishable within uncertainties.

\subsection{Analysis of volume-limited samples}

The approach adopted in our analysis does not completely eliminate the potential distortions introduced by observational selection effects.

This effect can be considered partially mitigated by examining subsamples constrained by a minimum redshift $z_{min}$. In each such subsample, the closest galaxies (which are relatively homogeneous) have the largest impact on determining the regression parameters, since the number of galaxies decreases significantly with increasing $z$. At the same time, the slope remains stable even under substantial changes in $z_{min}$. In the bin-based statistics --- which are less robust against observational selection but compensate for the enhanced influence of nearby galaxies --- the results remain consistent, demonstrating the robustness of the method.

To further reduce the influence of observational selection, we constructed volume-limited samples. These are subsamples of galaxies limited by a minimum brightness (i.e., a maximum absolute magnitude $M_{lim}$), such that all objects brighter than $M_{lim}$ would be observable up to a given redshift $z_{lim}$, and thus included in the sample. As the absolute magnitude, we used $M_{UV}$, computed according to \citet{oke1968energy, hogg2002k} as:
$$M_{UV}=m_{1500}-DM(z)-K(z),$$
where $m_{1500}$ is the apparent magnitude corresponding to rest-frame 1500\,\AA, $DM(z)$ is the distance modulus, and $K(z)$ is the $K$-correction. The value of $m_{1500}$ was computed using the catalogue's flux measurements, taking into account redshift. For the available photometric data, $M_{UV}$ could be reliably computed only for galaxies with $z \geq 2$. There are 2,659 such galaxies with reliable spectroscopic redshifts and 209,128 galaxies with photometric redshifts.

The distance modulus is given by: $$DM(z) = 5 \log_{10} \left( \frac{D_L(z)}{10\,\mathrm{pc}} \right),$$
where $D_L$ is the luminosity distance in the corresponding cosmological model \citep{hogg1999distance, lopez2010angular, laviolette2021expanding}:

$$D_L^{\Lambda{\rm CDM}}(z)=\frac{c}{H_0} \int\limits_0^z \frac{\mathrm dz'}{\sqrt{			\Omega_M (1+z')^3 + \Omega_\Lambda}}, 
$$

$$
D_L^{\rm TL}(z) = \frac{c}{H_0} \ln(1+z) \sqrt{1+z}.
$$

\begin{figure}
	\includegraphics[width=\hsize]{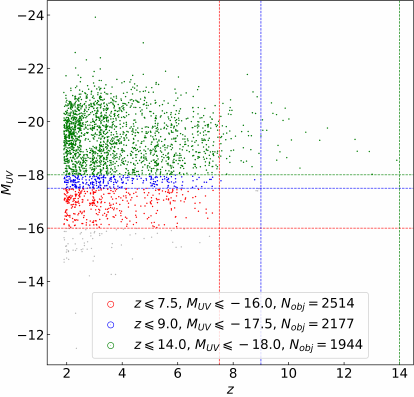}
	\includegraphics[width=\hsize]{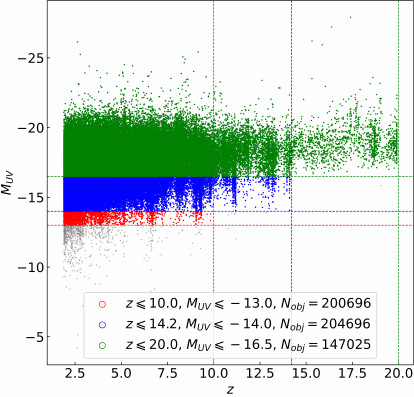}
	\caption{$z$ vs. $M_{UV}$ diagram and volume-limited samples in the $\Lambda$CDM model. Top panel: spectroscopic redshifts; bottom panel: photometric redshifts. Volume-limited samples are shown in colour.}
	\label{fig:zMlcdm}
\end{figure}

\begin{figure}
	\includegraphics[width=\hsize]{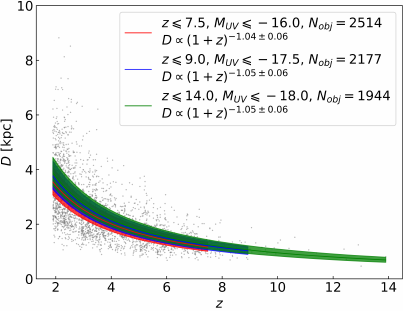}
	\includegraphics[width=\hsize]{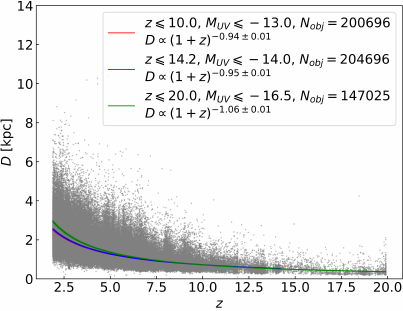}
	
	\caption{Best-fit curves for the evolution of galaxy sizes within the volume-limited samples shown in Fig.~\ref{fig:zMlcdm}, in the framework of the $\Lambda$CDM model. Top: spectroscopic redshifts; bottom: photometric redshifts.}
	\label{fig:vllcdm}
\end{figure}

The values of $M_{lim}$ and $z_{lim}$ were determined from the $\bigl(z, M\bigr)$ diagram. The corresponding diagrams and selected volume-limited samples are shown in Fig.~\ref{fig:zMlcdm}. A total of three volume-limited samples were used in each case. The analysis of the rate of evolution of galaxy sizes within these volume-limited samples was performed according to the methodology described in Section~\ref{rass}. The results are shown in Fig.~\ref{fig:vllcdm}, and the evolution slope coefficients are listed in Table~\ref{tab:vl}.

The obtained results are consistent with those presented earlier. However, the use of volume-limited samples reduces the formal uncertainty and makes the outcome more robust.

\begin{table}
\caption{Analysis of galaxy size evolution rate in volume-limited samples for two models. The table provides sample parameters (limiting absolute magnitude $M_{UV}$ and limiting redshift $z$), the number of galaxies in the sample, and the evolution slope $\alpha$ in the relation $D \propto (1 + z)^\alpha$, with its formal $3\sigma$ error.} \label{tab:vl}\vspace*{1em}

\begin{tabular}{c|cc|cc}
\hline
Model, $z$ & $z_{\lim}$ & $M_{lim}$ & $N_{obj}$ & $\alpha \pm 3\sigma$ \\
\hline
$\Lambda$CDM, Phot. & 10 & -13 & 200696 & $-0.94 \pm 0.01$ \\
$\Lambda$CDM, Phot. & 14.2 & -14 & 204696 & $-0.95 \pm 0.01$ \\
$\Lambda$CDM, Phot. & 20 & -16.5 & 147025 & $-1.06 \pm 0.01$ \\ \hline
$\Lambda$CDM, Spec. & 7.5 & -16 & 2514 & $-1.04 \pm 0.06$ \\
$\Lambda$CDM, Spec. & 9 & -17.5 & 2177 & $-1.05 \pm 0.06$ \\
$\Lambda$CDM, Spec. & 14 & -18 & 1944 & $-1.05 \pm 0.06$ \\
 
\end{tabular}

\end{table}

\section{Discussion and conclusion}


It is important to note that galaxies cannot serve as a standard ruler for two main reasons: their significant diversity in size and their evolutionary nature. The first issue can be mitigated by using a large sample and analysing general trends in the data. In this case, variations in characteristic galaxy sizes with redshift can be interpreted as the characteristic rate of their evolution within the cosmological model used to compute the linear size.

In this research, we analysed a statistically significant sample, which reveals a well-defined relationship between galaxy angular and linear sizes and redshift. Within the $\Lambda$CDM model, a pronounced rate of size evolution is observed, closely matching the rate of cosmic expansion. In contrast, the stationary ``tired light'' model shows no clear evidence of galaxy size evolution: the sizes of distant galaxies ($z>5$--10) are comparable to those of present-day giant galaxies.

The results of the regression analysis of the evolution of the sizes of galaxies support the authors' hypothesis that, within the standard $\Lambda$CDM cosmological model, the rate of evolution of galaxy sizes matches the rate of cosmic expansion. This evolutionary trend is observed up to at least $z \sim 10$ (for spectroscopic redshifts) and $z \sim 20$ (for photometric redshifts). This result appears paradoxical within the framework of the standard $\Lambda$CDM model, where gravitationally bound galaxies are not expected to expand due to the cosmic scale factor; any increase in their size should instead be driven by mechanisms unrelated to cosmology, such as star formation or accretion. This discrepancy warrants further investigation.

Our findings are qualitatively consistent with the results presented in \citet[Fig. 4]{allen2017size} and \citet[Fig. 5]{holwerda2020sizes}, based on pre-JWST observations, as well as with the more recent studies by \citet{ormerod2024epochs} and \citet{yang2025cosmos}, which rely on different galaxy samples, alternative methodologies, and do not explicitly compare cosmological models.

Resolving this apparent contradiction requires developing methods for estimating the linear sizes and masses of galaxies that are independent of the adopted cosmological model---for example, by using observed spectral line widths to estimate their gravitational potentials, as proposed in \citet{Raikov2024cosmological}.




\section*{Acknowledgements}
The authors are grateful to E.~Merlin for consultation on the ASTRODEEP-JWST catalogue data and to the anonymous referees for their valuable comments and helpful suggestions that significantly improved this work.

The ASTRODEEP catalogue used in this work is based in part on observations made with the NASA/ESA/CSA James Webb Space Telescope. The data were obtained from the Mikulski Archive for Space Telescopes at the Space Telescope Science Institute, which is operated by the Association of Universities for Research in Astronomy, Inc., under NASA contract NAS 5-03127 for JWST. These observations are associated with program JWST-ERS-1342. This catalogue is also based in part on observations made with the NASA/ESA Hubble Space Telescope obtained from the Space Telescope Science Institute, which is operated by the Association of Universities for Research in Astronomy, Inc., under NASA contract NAS 5-26555. These observations are associated with program HST-GO-17321. 

\section*{Funding}
A.~A.~Raikov carried out this work within the framework of the state assignment of the Special Astrophysical Observatory of the Russian Academy of Sciences, approved by the Ministry of Science and Higher Education of the Russian Federation. The other co-authors received no additional funding.

\bibliographystyle{abbrvnat}
\bibliography{Test-1}

\end{document}